\documentclass[journal]{IEEEtran}

\usepackage{cite}
\usepackage{color}
\usepackage{graphicx,epsfig}
\usepackage{epstopdf}
\usepackage{caption}
\usepackage{amssymb}
\usepackage[cmex10]{amsmath}

\begin{document}
%
\title{Map-based Millimeter-Wave Channel Models: \\An Overview, Hybrid Modeling, Data, and Learning} 

\author{Yeon-Geun~Lim,~\IEEEmembership{Student~Member,~IEEE,}
        Yae~Jee~Cho,~\IEEEmembership{Student~Member,~IEEE,} 
        Min~Soo~Sim,~\IEEEmembership{Student~Member,~IEEE,}
        Younsun~Kim,~\IEEEmembership{Member,~IEEE,}\\
        Chan-Byoung~Chae,~\IEEEmembership{Senior~Member,~IEEE,} and 
        Reinaldo A. Valenzuela,~\IEEEmembership{Fellow,~IEEE}
\thanks{Y.-G. Lim, M.~S.~Sim, and C.-B.~Chae are with the School of Integrated Technology, Institute of Convergence Technology, Yonsei University, Korea (e-mail: \{yglim, simms\}@yonsei.ac.kr, cbchae@ieee.org). Y.-J.~Cho is with Mercedes-Benz  Research \& Development, Korea (e-mail: yaejee.cho@daimler.com). Y.~Kim is with Samsung Electronics, Korea (e-mail: younsun@samsung.com). R. A. Valenzuela is with Nokia Bell Labs., NJ, USA (e-mail: reinaldo.valenzuela@nokia-bell-labs.com). Corresponding author is {C.-B. Chae.}}
\thanks{This research was supported by the MSIT (Ministry of Science and ICT), Korea, under the ``ICT Consilience Creative Program'' (IITP-2019-2017-0-01015) supervised by the IITP (Institute for Information \& communications Technology  Promotion).}}


\maketitle

\begin{abstract}

Compared to the current wireless communication systems, millimeter wave (mm-Wave) promises a wide range of spectrum.
As viable alternatives to existing mm-Wave channel models, various map-based channel models with different modeling methods have been widely discussed. Map-based channel models are based on a ray-tracing algorithm and include realistic channel parameters in a given map. Such parameters enable researchers to accurately evaluate novel technologies in the mm-Wave range. Diverse map-based modeling methods result in different modeling objectives, including the characteristics of channel parameters and different complexities of the modeling procedure. This article outlines an overview of map-based mm-Wave channel models and proposes a concept of how they can be utilized to integrate a hardware testbed/sounder with a software testbed/sounder. In addition, we categorize map-based channel parameters and provide guidelines for hybrid modeling. 
Next, we share the measurement data and the map-based channel parameters with the public. Lastly, we evaluate a machine learning-based beam selection algorithm through the shared database. We expect that the offered guidelines and the shared database will enable researchers to readily design a map-based channel model.

\end{abstract}


%
\IEEEpeerreviewmaketitle


\section{Introduction}\label{Section:Intro} 

\IEEEPARstart{F}{uture} fifth-generation (5G) communication systems will include substantial types of service involving enhanced mobile broadband, massive machine-type communication (mMTC), and ultra-reliable low-latency communication. Researchers have discussed numerous applications of such service types, including the Internet of Things, Internet of drones, gigabit wireless connectivity, and autonomous vehicles. They have been also debating potential frequency bands to service such applications~\cite{Lim_WCM18}. These new applications and services will be launched in the millimeter-wave (mm-Wave) range due to a shortage of bandwidth in the sub-6 GHz bands, which legacy radio communication systems have made tremendous use of because of the excellent radio propagation property. Mm-Wave channel models have been thoroughly investigated through extensive measurements and simulations and accurate models are vital for the design of mm-Wave communication systems.

The sub-6 GHz channel models for system-level evaluation, which are usually based on geometry-based stochastic channel models (GSCMs), have focused on evaluating the performance of point-to-point communications or multiple-input multiple-output (MIMO) systems with a small number of antennas (up to eight in LTE). These models have  performed with the regular cell size of LTE (e.g., 200~m for an urban micro (UMi) scenario) and are based on a base station to user equipment (BS-UE) link type. Recent mm-Wave channel models have not only included the inherent characteristics of the mm-Wave channels, but have also added channel properties for 5G communication technologies, such as massive MIMO (M-MIMO) and hybrid beamforming~\cite{SSCM_TMTT, METISmodel, 3GPP_mmwave}. Nevertheless, these existing GSCMs are concentrated on modeling BS-UE links with a regular cell size and cannot support all modeling requirements for new 5G applications due to the lack of channel measurement campaigns for various link types~\cite{ComMag_METIS}. 

Map-based mm-Wave channel models that utilize ray-tracing (RT) have gathered momentum. These models serve as a way to model the irregular layouts of small cells and to support new applications' link types, including device-to-device (D2D), vehicle-to-everything (V2X), and air-to-everything (A2X). Moreover, researchers have utilized RT not only to evaluate hardware (HW) testbeds~\cite{Kwon_TMTT, YJ_ComMag18,RTmmwave_JSAC} but also to validate the theoretical performance of vital 5G technologies~\cite{Lim_WCM18}, since RT can cover a wide range of modeling specifications and has only a minor discrepancy with HW measurements~\cite{Rappaport_SiteSpecific,ComMag_METIS,RTmmwave_JSAC,WiSE_JSAC}. 
Some popular mm-Wave channel models have also adopted RT techniques or map-based models. For example, 
NYU WIRELESS developed the NYUSIM model by complementing HW channel measurements with RT~\cite{SSCM_TMTT}. The METIS group proposed a map-based channel model supporting various modeling requirements of 5G~\cite{METISmodel}. The 3GPP model adopted a hybrid of a GSCM and a map-based model~\cite{3GPP_mmwave}, which was accepted in the ITU-R IMT-2020 evaluation report.

\begin{figure*}[!t]
	\centerline{\resizebox{2\columnwidth}{!}{\includegraphics{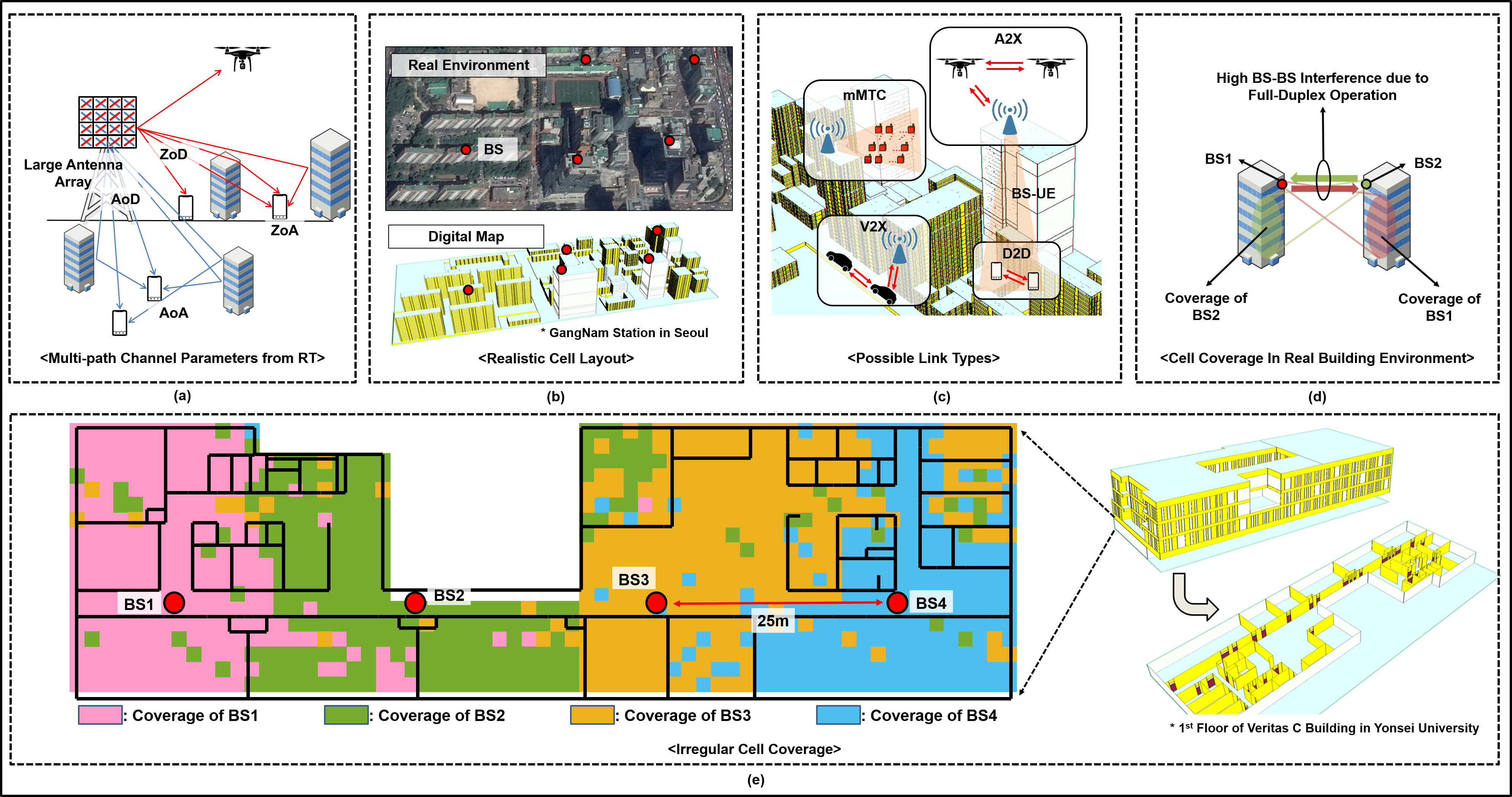}}}
	\caption{An illustration of a map-based model and its characteristics: (a) multipath channel parameters from RT; (b) a realistic cell layout in the digital map of GangNam Station, Seoul, South Korea; (c) possible link types; (d) cell coverage in real building environment. In this case, BS-to-BS interference is non-negligible; (e) irregular cell coverages based on a digital map of Veritas C in Yonsei University. In this illustration, WiSE, an RT software developed by Bell Labs~\cite{WiSE_JSAC}, is used to show the digital map and predict coverage. The heights of the BSs and the UEs are, respectively, 3~m and 1~m, and the center frequency is 28~GHz.}
	\label{Figure:Map_based_Model}
\end{figure*}

Before adopting the hybrid channel model, the engineers behind the 3GPP model discussed many kinds of hybrid channel modeling methods that differed in both generating and combining the channel parameters of a GSCM and a map-based model. Since such methods not only result in different modeling objectives, including characteristics of the channel parameters and complexities of the modeling procedure, but also have the potential to be used in different venues, how to appropriately categorize these methods remains an open area for study at 3GPP to make the modeling objective of each hybrid approach clear. 
Note that the conventional classifications of map-based channel models have not specifically categorized hybrid models, although they have included two categories, a deterministic (and/or semi-deterministic) channel model and a hybrid channel model between the first category and a GSCM. 

Recently, researchers have utilized map-based channels to evaluate the accuracy of machine learning-based algorithms and to consider practical scenarios~\cite{Klautau_18ITA}. Machine learning-based algorithms are used for searching a pattern of certain parameters. Training data for evaluating this algorithm should have user- or cell-specific information, which means map-based channel models are attractive candidates of training data.

This article provides an overview of map-based mm-Wave channel models. The motivation for it lies in the fact that researchers have paid attention to designing a map-based mm-Wave channel model, while still considering why map-based channel models--rather than GSCMs--should be utilized in the mm-Wave range. We also propose a categorization of map-based channel parameters and provide guidelines for modeling each categorization. Finally, we share with the public the measurement database of the channel parameters for specific indoor and outdoor regions and for evaluating machine learning algorithms, which anyone can download and use for free.\footnote{The details of the database and its manual are available at http://www.cbchae.org/. We also present a demo video of our RT simulations in various scenarios, including the digital map realization methods.}

To the best of our knowledge, this is the first work that not only categorizes map-based hybrid channel models but also shares the database of map-based channel parameters. This categorization ensures that a hybrid channel model is efficiently utilized by considering the different modeling objectives and requirements for various 5G mm-Wave systems.
Most prior work on map-based channel models~\cite{Rappaport_SiteSpecific,METISmodel} were focused on proposing RT algorithms, offering specific procedures for channel coefficient generation, and validating their models. 
Nonetheless, some researchers might not be familiar with modeling the map-based channel or realizing a digital map that depicts real-world buildings to be used for RT. We expect that both the offered guidelines and the shared database will encourage researchers, even those researchers without RT software, to readily design a map-based channel.

\section{Why Should We Utilize Map-based Channel Models in the mm-Wave Range?}\label{Section:why}

In this section, we start with a brief update on the recent progress of existing mm-Wave GSCMs. Then, we provide the advantages of map-based channel models compared to GSCMs.

Mm-Wave is going to be an important component of 5G deployment.
However, some inherent properties of mm-Wave propagation-links, such as high path loss, high penetration loss, and blockages, induce high propagation-link loss. Consequently, engineers must try to achieve a higher propagation gain using technologies such as large array antenna systems. 
In such systems, the spherical wave assumption (not plane wave assumption) is plausible due to the large size of the arrays regarding wavelength. Moreover, if receivers (RXs) are located a short distance away from each other in a multi-user M-MIMO or an mMTC scenario, the correlations of cluster- and ray-specific random variables are very high; that is, the channel is spatially consistent~\cite{ComMag_METIS, Ju18GC}. 
Considering these characteristics of the mm-Wave channel, the additional features in the existing mm-Wave GSCMs can be summarized as follows:\footnote{We concentrate on some popular mm-Wave channel models, such as the 3GPP model, the METIS model, and the NYUSIM model in this article.}
\begin{itemize}
	\item {Channel parameters, such as large-scale parameters (LSPs--e.g., path loss, delay spread (DS), angle spread (AS), the number of clusters and rays, etc.) and small-scale parameters (SSPs--e.g., azimuth angle of departure (AoD), azimuth angle of arrival (AoA), zenith angle of departure (ZoD), zenith angle of arrival (ZoA), power delay profile (PDP), etc.), are determined by considering the center frequencies of various operating bands and frequency selectivity due to the broad system bandwidth.}
	\item {The received powers and the total number of clusters and rays are relatively fewer than those in the sub-6~GHz range by taking into account high propagation-link loss.}
	\item {The spatial consistency can be applied based on the correlation distance and user's movement.}
	\item {With the large array antenna assumption, one can model the individual time of arrival and angle offsets for all rays per link between transmit and receive antennas.}
\end{itemize}

\begin{figure*}[!t]
	\centerline{\resizebox{1.9\columnwidth}{!}{\includegraphics{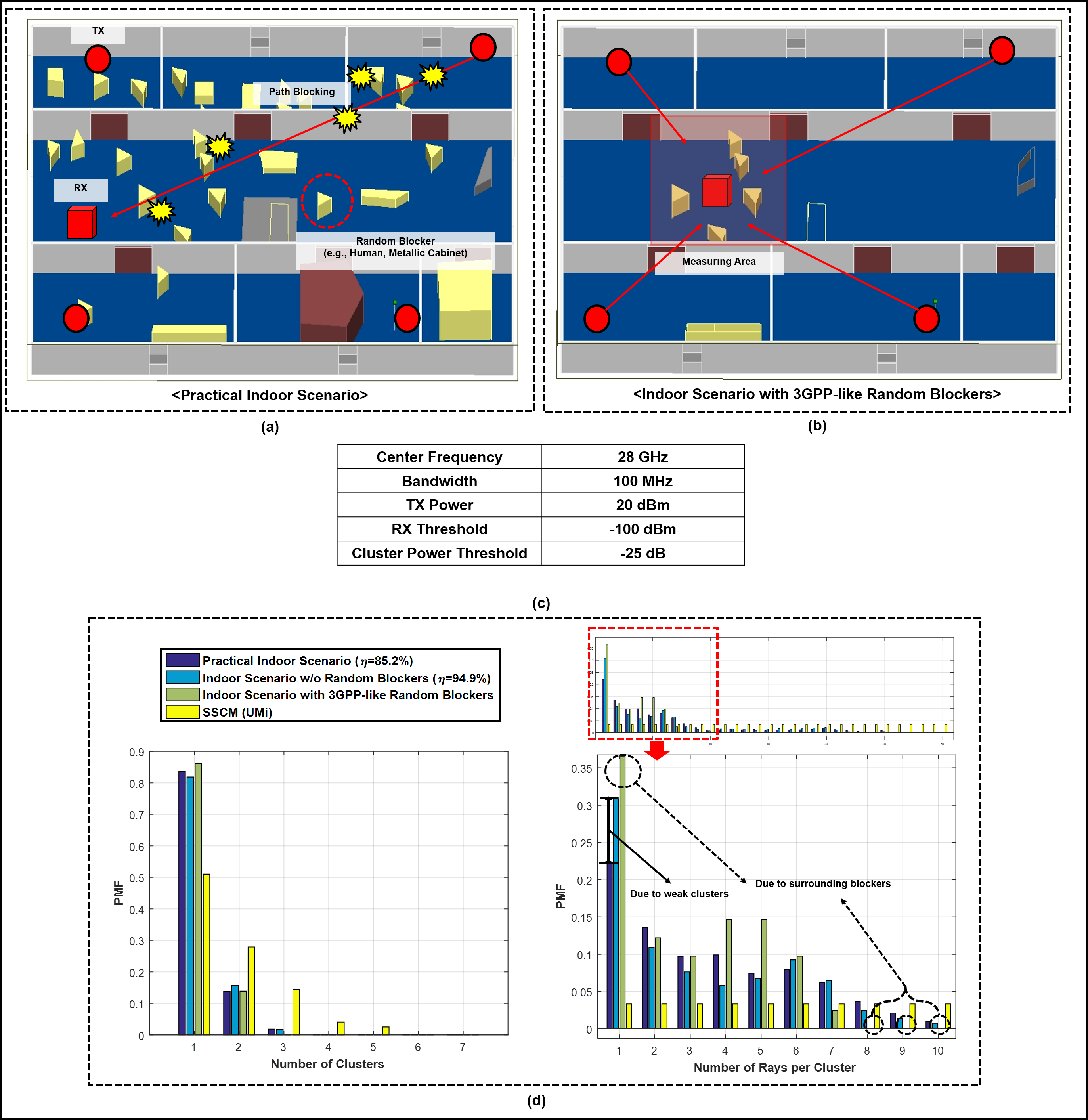}}}
	\caption{Channel parameter modeling using map-based channels: (a) the digital map for a practical indoor scenario where the RXs are uniformly distributed in a building; (b) an indoor scenario with 3GPP-like random blockers where channel parameters are measured in the shaded area; (c) simulation parameters; and (d) simulation results. In this illustration, we use Wireless InSite, the RT software developed by Remcom. The coverage is denoted as $\eta$.}
	\label{Figiure:LLS}
\end{figure*}

\subsection{New Modeling Requirements of Emerging Applications in the mm-Wave Range} 
Map-based channel models (also known as site-specific propagation models) generate multi-path channel parameters, as illustrated in Fig.~\ref{Figure:Map_based_Model}a, by utilizing RT~\cite{Rappaport_SiteSpecific,METISmodel}. A customized three-dimensional digital map is used to organize a realistic cell layout, as shown in Fig.~\ref{Figure:Map_based_Model}b. Although the existing mm-Wave GSCMs have exploited the essential features of the mm-Wave channel as mentioned above, these features and their target scenarios are limited by a link type such as traditional BS-UE as listed in Table~I. For example, it is hard to design outdoor channels with different inter-site distances (ISDs) shorter than 200~m because the statistical channel parameters of GSCMs is valid within their minimum ISD, 200~m, and with regular cell topologies. Many types of applications, however, will appear with different link types as shown in Fig.~\ref{Figure:Map_based_Model}c. \emph{Note that covering all modeling requirements for such applications is almost impossible with existing GSCMs, which still have insufficient channel measurement campaigns.} Meanwhile, in addition to including the essential features of mm-Wave channel in the GSCMs, map-based models also handle the following modeling requirements that are not limited:

\emph{\bf Short-Range Communication Links:} The shorter the link is, the more the map-based models can depict the channel characteristics influenced by surrounding topography.
In 5G, cell sizes are likely to shrink for high network density. In practical small cell topographies, the coverage of each cell varies due to the assorted shapes and heights of surrounding structures, as illustrated by the irregular cell coverages in Figs.~\ref{Figure:Map_based_Model}d and~\ref{Figure:Map_based_Model}e. Thus, the level of the intercell interference will be dependent on the real geometry while in a GSCM, this level is mainly dependent on the distance between the BS and the UE due to the regular cell layout. Fig.~\ref{Figure:Map_based_Model}d shows that the level of the BS-to-BS interference can be high as much as the received power at downlink in the full-duplex operation. This level was underestimated in previous studies with GSCMs due to the large separation between the BSs.

\emph{\bf Realistic Indoor Environments:}
In outdoor-to-indoor (O2I) paths, if RXs are located indoors, penetration loss by both external and internal walls should be considered for more accurate modeling. Although many scenarios support indoor users (e.g., 80~percent of users in outdoor scenarios) and penetration loss is essential in a mm-Wave channel, the 3GPP model does not fully consider internal walls. In indoor-to-indoor (I2I) paths, blockers can be in the middle of the paths if a transmitter (TX) and an RX are at similar heights.

\emph{\bf Various Mobility Types:}
The new applications of mm-Wave systems will support various mobility types. Both dual-mobility of the link ends and the mobility of blockers should also be considered to ensure accurate channel-modeling for certain link types such as the D2D link. Notably, in indoor communications, researchers should take into consideration devices moving from room to room or from floor to floor.

\emph{\bf User-Specific Channel Parameters:}
In GSCMs, LSPs of each cell are randomly generated from the same distribution, regardless of cell topography; SSPs per user are generated from the cell-specific LSPs, regardless of users' location within the cell (except for LOS angle parameters). These would yield an inaccurate performance evaluation of cell- and user-specific technologies. These technologies include, for instance, the beam selection based on machine learning that classifies beam indices based on user-specific information~\cite{Klautau_18ITA}. 

\emph{\bf Channels on Different Frequencies:}
Map-based channel models can measure channels of a specific TX/RX pair on different frequencies, while GSCM cannot measure them.
5G will allow the devices of which functions are operated on different frequencies. These devices include numerology multiplexing, which supports different service types on different subbands~\cite{Lim_WCM18}, integrated access backhual technologies, and the beamforming operation when control and data signals are on different frequency bands.


\subsection{Consistent Evaluation for the Target Scenario}

Recent mm-Wave technologies have emerged within various scenarios. Some technologies can enhance system performance by utilizing characteristics of the mm-Wave channel. To prevent either over- or underestimated link-level evaluation, researchers should use an appropriate channel model that depicts well the servicing scenario of their proposed technologies.
For example, Fig.~\ref{Figiure:LLS}d shows the probability mass functions (PMFs) of the number of the clusters and rays that are measured from an indoor digital map in~Figs. \ref{Figiure:LLS}a and \ref{Figiure:LLS}b. To obtain the PMFs, three scenarios are considered--a practical indoor scenario in which there are many blockers in the building, an indoor scenario with no blockers, and an indoor scenario with five blockers around the RX (similar to the 3GPP blockage model). 
Figure~\ref{Figiure:LLS}d shows similarity in the PMFs of the number of the clusters among three indoor scenarios, and the number of clusters can be smaller than that of an UMi scenario derived from the distribution in~\cite{SSCM_TMTT}. For practical purposes, the coverage shrinks, and some weak clusters consisting of one ray vanish due to the presence of many blockers. The maximum number of rays per cluster decreases when deploying 3GPP-like blockers due to a good deal of surrounding blockers. These results are related to the rank of a channel, which affects not only MIMO performance, but also the accuracy of channel estimation algorithms that utilize the sparsity of the channel.

\subsection{Map-based SW Testbed/Sounder}\label{Section:Integ.HW}

One significant use case of the map-based channel model is the integration of a HW and a software (SW) testbed/sounder. This use case has not  been discussed widely.
We can reconfirm the result of the HW testbed by link-level evaluation in the map-based model that describes the test site of the HW testbed; and we can extend this evaluation into system-level evaluation of a HW testbed. Besides, the real-world channel measurements can be double-checked in a similar way.

\emph{\bf Feasible System-level Evaluation of a HW Testbed Using a Map-based SW Testbed:}
In the interest of the development of novel mm-Wave technologies, both their theoretical model and algorithm implementation have been jointly evaluated, under various scenarios, through a SW testbed from the link to system levels. Notwithstanding the versatility of a SW testbed, it is crucial to prototype a HW testbed before implementing technologies in the real world. 
System-level evaluation using a HW testbed is, however, laborious, so a HW testbed usually assesses a technology at the link-level. System-level evaluation is desirable due to the complex radio-links of the mm-Wave systems supporting complicated scenarios for various applications. If system-level evaluation using a HW testbed is practically feasible, it will promote advancing the mm-Wave system.

One option to perform system-level evaluation using a HW testbed is to integrate the HW testbed with a map-based SW testbed. With reflecting a link-level evaluation result of a HW testbed into that of a SW testbed (i.e., calibration), the system-level evaluation can be conducted in the digital map consisting of the test site. We describe two related work below.
\begin{itemize}
	\item The authors in~\cite{Kwon_TMTT} and~\cite{YJ_ComMag18} fabricated RF lenses that operated at 77~GHz and 28~GHz respectively, and measured their HW performance. They then evaluated an RF lens-embedded MIMO system at the system level by combining HW measurements and specific algorithms, which were a proposed codebook in~\cite{Kwon_TMTT} and the standard mm-Wave beamforming scheme in 802.11 (exhaustive search by beam sweeping) in~\cite{YJ_ComMag18}, with map-based channels.
\end{itemize}

\begin{figure*}[!t]
	\centerline{\resizebox{2\columnwidth}{!}{\includegraphics{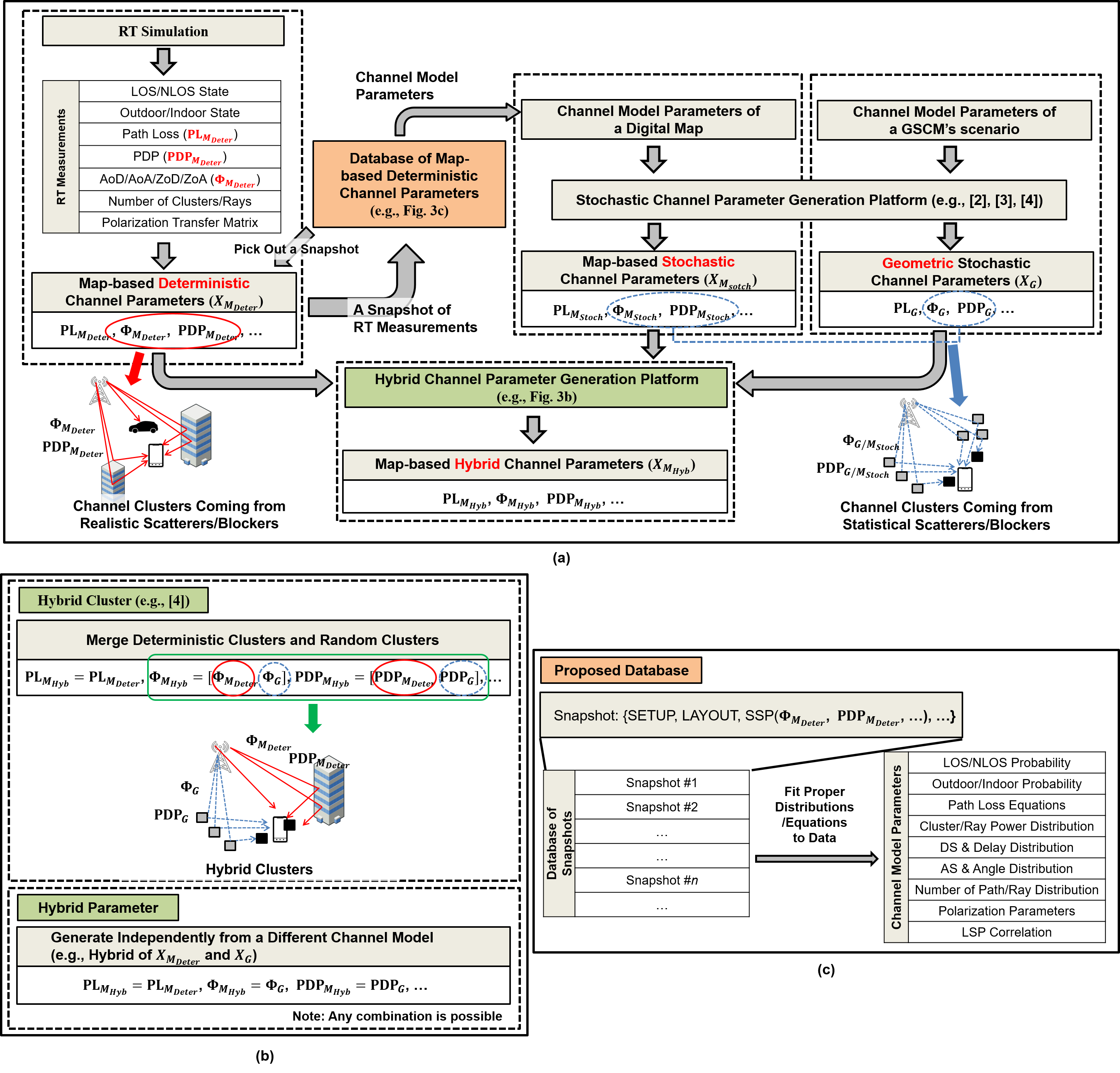}}}
	\caption{An illustration of categorizations and generalized channel parameter generation: (a) a block diagram of channel parameter generation for each category; (b) subcategories and examples of map-based hybrid channel parameters; (c) an illustration of the proposed database.} 
	\label{Figure:BD}
\end{figure*}

\emph{\bf Channel Measurement based on a Map-based SW channel sounder:}
Another use of the integration based on a map-based channel model is to allow it to play the role of a SW channel sounder by utilizing the measurements of a real-world channel sounder. This concept enables the measurement of channel characteristics and the assessment of theoretical technologies, and it includes the existing mm-Wave channel model as follows:
\begin{itemize}
	\item The authors in~\cite{RTmmwave_Access} measured LSPs and SSPs by using a 60~GHz channel sounder. They calibrated the map-based channel and evaluated theoretical beamforming technologies in a digital map.
	\item {NYU WIRELESS utilized RT results to complement the HW channel sounder. RT recreated the absolute propagation time of arrivals from BS-UE links and retrieved AoA distribution for the validity of the model~\cite{SSCM_TMTT}.}
\end{itemize}

\begin{table*}[t!]
	\caption{Summary of the Proposed Database and the Conventional mm-Wave Channel Models}
	\centerline{\resizebox{2\columnwidth}{!}{\includegraphics{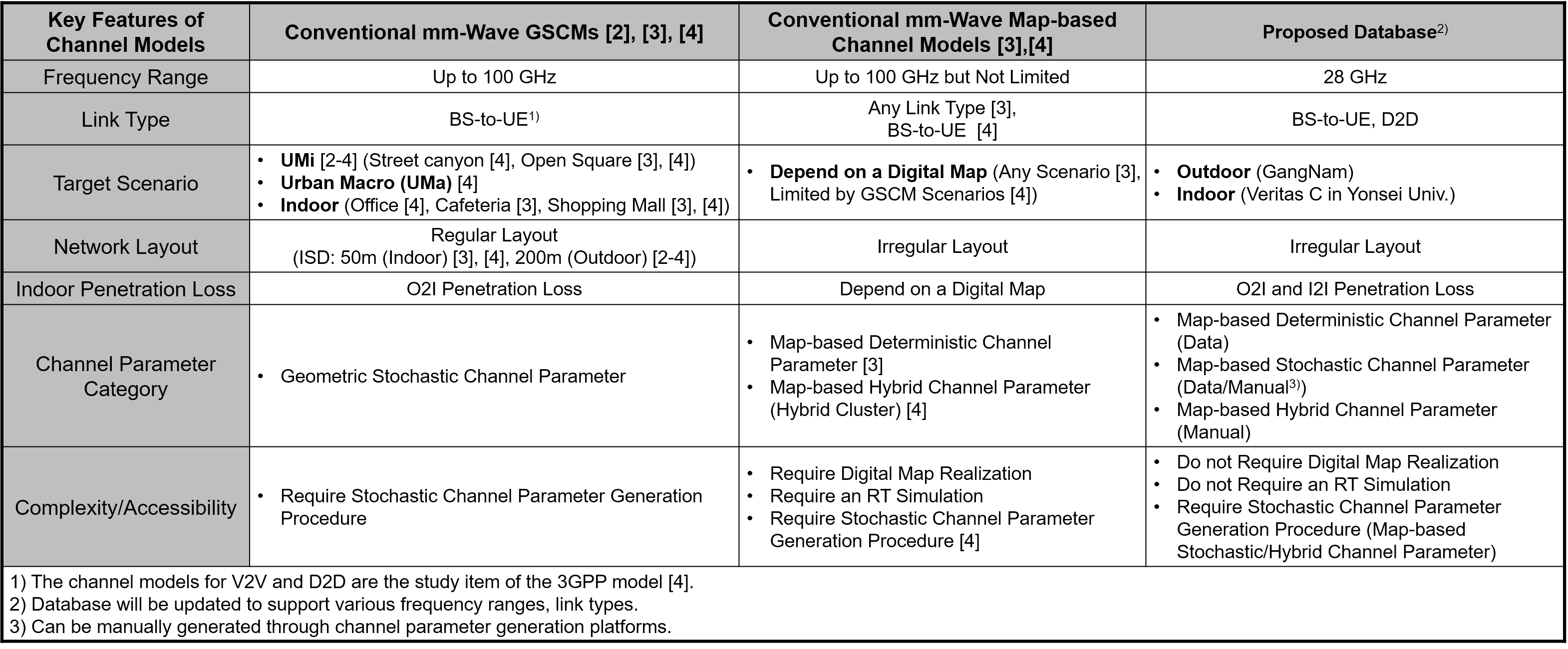}}}\vspace{2mm}
	\label{Table1}
\end{table*}

\begin{figure*}[!t]
	\centerline{\resizebox{1.6\columnwidth}{!}{\includegraphics{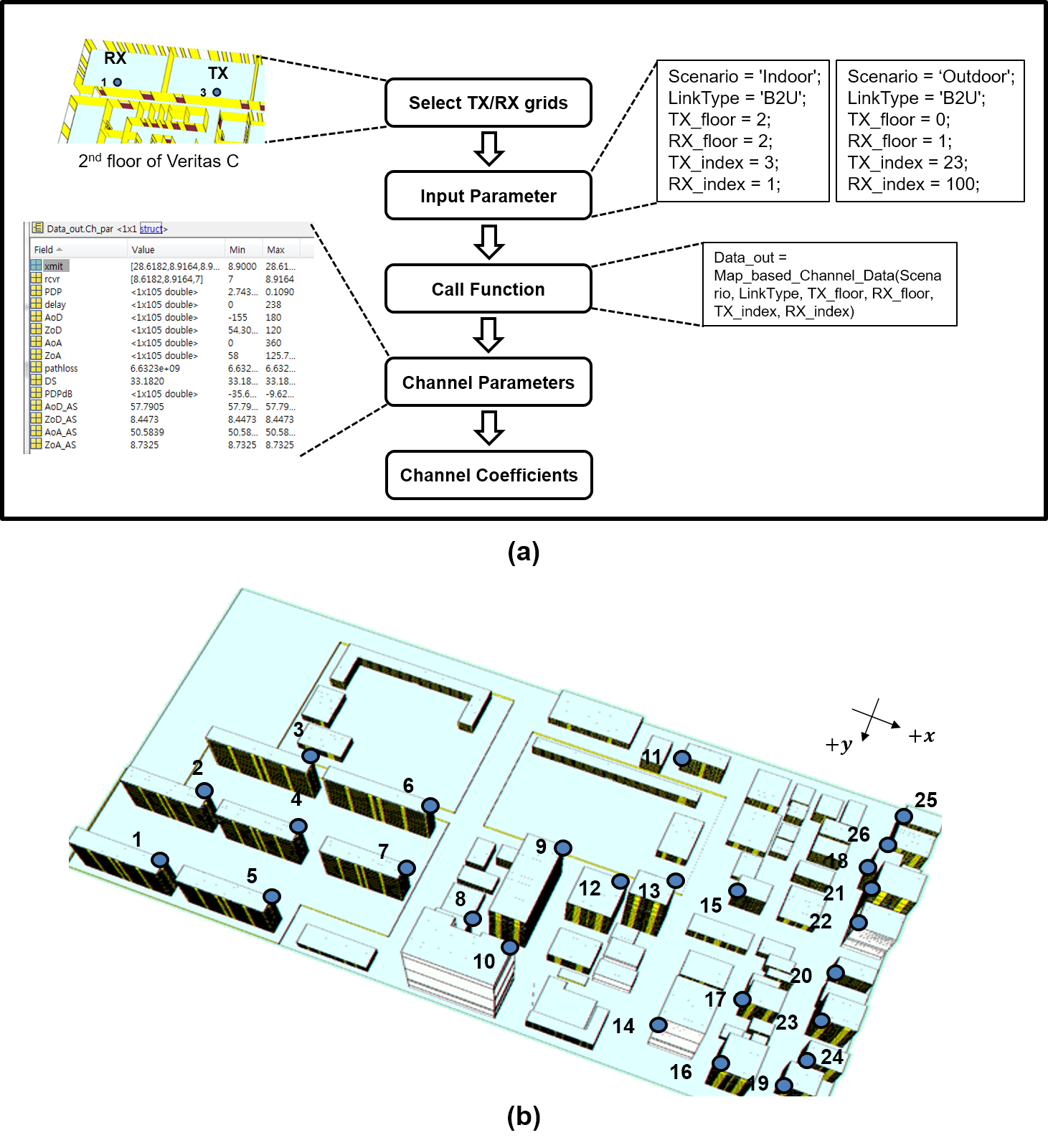}}}
	\caption{An illustration of the proposed database: (a) an example procedure for channel coefficient generation using the proposed database; (b) an example layout of grids. In this illustration, the grids of the BS locations in the outdoor scenario are shown.} 
	\label{Figure:database}
\end{figure*}

\section{Categorization and Guidelines for Generating Map-based Channel Parameters}

In this section, we categorize map-based channel parameters from a modeling approach perspective. Then, we provide guidelines for both modeling each category and selecting a proper category. Figure~\ref{Figure:BD} shows an illustration of the categorizations and generalized channel parameter generation. Before generating channel parameters, a user first selects a proper channel model according to the target scenario and the modeling requirements. In the next step, the user determines the system setup and a network layout. We describe this step if a map-based channel model is selected as follows:
\begin{itemize}
	\item {\emph{System Setup:} In the first step, the digital map is realized based on material information of walls provided in~\cite{ITU_material}.\footnote{The measurement results for building materials in various center frequencies are available at http://wireless.engineering.nyu.edu/ (e.g., \cite{Xing18GC}).}
	This map represents the area and scenario for the target application and its adopted technologies. The second step involves setting up the system parameters of the target application's TX/RX such as antenna patterns of the TX/RX, their array configurations, and so on.
		
		
	} 	
	\item {\emph{Network Layout:} TXs/RXs as well as random objects are dropped into the digital map. Next, a beamforming pattern is applied together with the resource allocation for each TX/RX combinations.}
\end{itemize}

Finally, channel parameters such as LOS state and SSPs, are generated. Note that details of channel parameter generation methods are dependent on the RT simulator and/or the platform such as--the stochastic channel parameter-generating platforms in \cite{SSCM_TMTT,METISmodel,3GPP_mmwave}--which can be chosen by a user.
As argued in the following subsections, these channel parameters can be categorized as shown in Fig.~\ref{Figure:BD}a.

\subsection{Category I -- Map-Based Deterministic Channel Parameters} 

An RT simulator generates deterministic channel parameters at the particular TX/RX locations. This approach ensures that the channel parameters are accurate for a given network layout. Since the channel snapshot represents only the characteristics of the given locations of the random scatterers and blockers, their locations change according to their mobility and trajectory at the next snapshot; this is contrasted with stochastic parameters in which their impact is involved. The channel parameters generated from either the conventional map-based channel models~\cite{METISmodel} or the typical RT simulators fall into this category. 

\subsection{Category II -- Map-Based Stochastic Channel Parameters}
The channel model parameters such as stochastic LSPs are determined from as many deterministic channel snapshots as possible in both a target scenario and a given digital map. Map-based stochastic channel parameters are generated according to the chosen stochastic channel parameter-generating platform by replacing \emph{the chosen platform's channel model parameters} with \emph{the fitted channel model parameters of the snapshots}. Within the mm-Wave range, this category has yet to be studied. Meanwhile, in the sub-6 GHz range, the channel model in~\cite{Lim_WCM18} falls into this category where the authors fitted LSPs to their RT measurements and generated channel parameters through the platform in~the 3GPP model.

\subsection{Category III -- Map-Based Hybrid Channel Parameters}
Hybrid channel parameters are complementary to the channel parameters from a different channel model or to the HW measurements. We present modeling guidelines for three subcategories of map-based hybrid channel parameters. The examples are described in Fig.~\ref{Figure:BD}b
 
\emph{\bf Category III.A -- Hybrid Cluster:}
 The channel parameters consist of deterministic clusters that are composed of deterministic channel parameters and random clusters that are composed of GSCM's channel parameters. These clusters are treated as independent clusters, but are merged after clustering. For example, the map-based deterministic channel parameters for deterministic clustering and the GSCM's channel parameters for random clustering merge into the hybrid cluster, which means both angle parameters and PDPs are generated from both models. The 3GPP model supports this subcategory because of good compatibility with its own GSCM.

\emph{\bf Category III.B -- Hybrid Parameter:} To compensate for the shortcomings of each channel parameter, each one is independently generated from a different channel model. For example, PDPs are generated from a map-based stochastic channel model and path losses are generated from a map-based deterministic channel model. The advantage of hybrid parameter compared with the hybrid cluster is the simpler implementation it permits. 

\emph{\bf Category III.C -- Hybrid of Deterministic Parameters and HW Measurements:} In this case, the map-based deterministic channel parameters are combined into the channel coefficient with the calibration factors, as mentioned in Section~\ref{Section:Integ.HW}. The NYUSIM model utilized this subcategory.

\subsection{Guidelines for Category Selection}

Here, we provide some guidelines for selecting a category regarding the target link type/scenario and modeling requirements. We assume that, as is the case with GSCMs, the channel model parameters of a digital map are predetermined.

\begin{itemize}
	\item {\emph{Category I}, \emph{Category II}, or \emph{the hybrid of both} should be selected if channel parameters need to almost perfectly reflect practical building environments.}
	\item { \emph{Category III.A} or \emph{Category III.B between a GSCM and a map-based channel model} can be selected if channel parameters need to reflect both practical building environments with irregular cell layouts and the statistical real-world channel measurements of the target scenario. A recommendable use case is the 3GPP hybrid model that generates random clusters by combining GSCM's parameters and deterministic clusters' parameters for their compatibility~\cite{3GPP_mmwave}. Another use case is that the user-specific channel parameters are predicted from a map-based channel model while the others are generated through a GSCM.}
	\item {If \emph{Categories I-III} can be selected, the selection among them depends on modeling complexity. The details are given below.}
\end{itemize}

The higher accuracy \emph{Category I}'s parameters have, the higher their generating complexity gets. Each snapshot of \emph{Category I} represents the deterministic channel for a given TX/RX location while \emph{Category II}'s parameters are not fully dependent on a TX/RX location. 
Generating \emph{Category II}'s parameters has equal complexity with the GSCMs in the same generation platform while these parameters still represent the characteristic of the digital map. It is thus obvious that the complexity of generating \emph{Category III}'s parameters are intermediate between them.

\begin{figure*}[!t]
	\centerline{\resizebox{2\columnwidth}{!}{\includegraphics{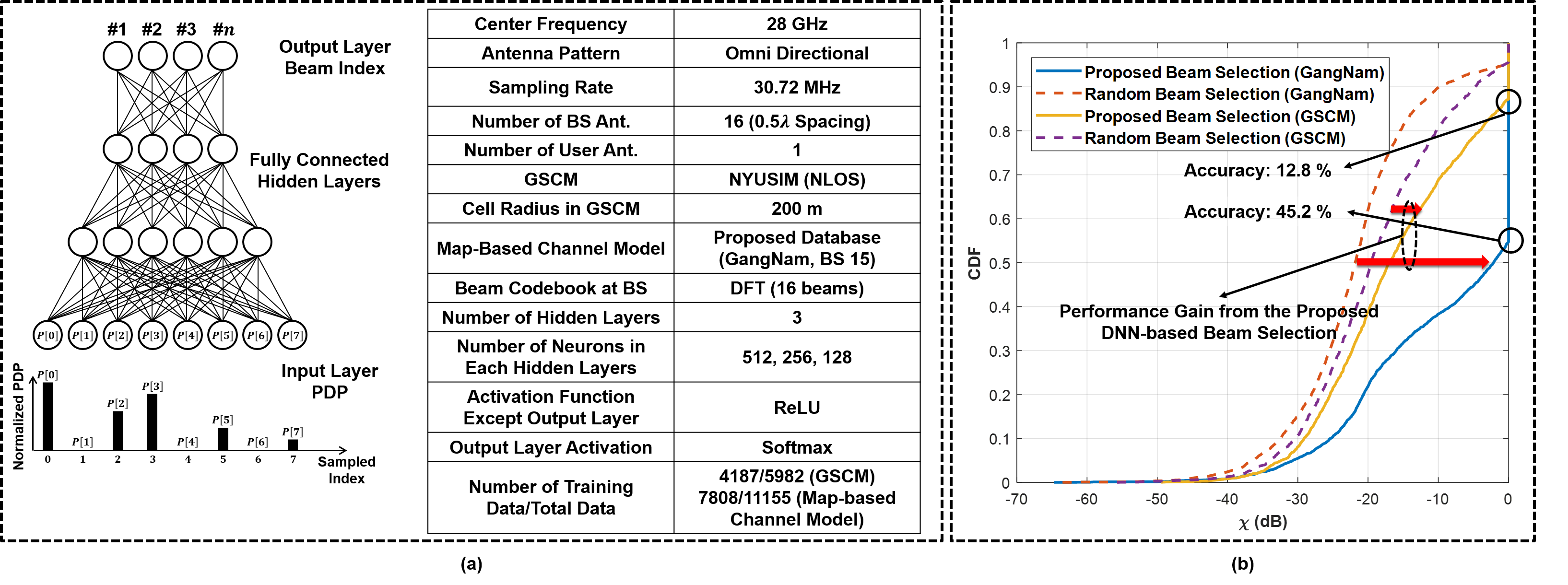}}}
	\caption{Performance evaluation through the proposed database: (a) an illustration of the proposed DNN-based beam selection algorithm and the simulation parameters; (b) CDFs of the performance of the proposed algorithm and random beam selection with different channel models. The x-axis represents the received power gain from algorithms against the exhaustive search by beam sweeping.} 
	\label{Figure:DNNresults}
\end{figure*}


\section{Database of mm-Wave Map-based Channel Parameters}

\subsection{Concept for a Database of Deterministic Channel Snapshots} A database of deterministic channel snapshots has been devised to reduce the complexity of RT. Figure~\ref{Figure:BD}c illustrates that a channel snapshot can be picked out from the database according to the specific TX/RX locations for a given system setup. This method not only maintains the accuracy of the channel, but also reduces to a tremendous degree the complexity of channel generation (because it only requires a picking-out-algorithm). In addition, the database is necessary to determine stochastic LSPs for \emph{Category II}.

\subsection{Introduction to the Proposed Database}
In this section, we briefly introduce the proposed database. The database targets specific indoor and outdoor scenarios that were measured from the manufactured digital maps of the Veritas~C building in Yonsei University and Gangnam Station in Seoul, respectively, supporting the BS-UE and D2D link types. Table I summarizes the key feature of the channel model of the proposed database compared with the conventional mm-Wave GSCMs and map-based channel models. The main advantages of the proposed database are that any researcher can design map-based channel models without an RT simulator and that digital map realization is not required.

Figure~\ref{Figure:database}a illustrates an example procedure for channel parameter generation through the proposed database. We provide a massive number of snapshots of TX/RX pairs. Thus, a user simply selects TX/RX grids and inputs a few parameters, including the scenario, the link type, and the TX/RX's floors. For example, a user can drop a TX in the digital map of Gangnam station by selecting a BS grid in Fig.~\ref{Figure:database}b. Then, the database will provide the deterministic channel parameters of the selected TX/RX grids and the user-specific stochastic LSPs, DS and AS. The channel coefficient can be readily generated from the typical clustered channel model or through a chosen channel parameter generation platform.

\subsection{Training a Machine Learning Algorithm with the Proposed Database}

	The proposed database can provide a massive number of snapshots for a machine learning training set. Through this training set, we will show why map-based channels are more appropriate for evaluating machine-learning-based algorithms. We first propose a deep neural network (DNN)-based beam selection algorithm as illustrated in Fig.~\ref{Figure:DNNresults}a. Its input is a PDP the number of rays of which is two or more and its output is the index of a beam codebook. Similar to the algorithm in~\cite{Klautau_18ITA}, the required time for beam selection can be quite reduced compared with the exhaustive search by beam sweeping. 
	Figure~\ref{Figure:DNNresults}b shows CDFs of the performances of the proposed algorithm and random beam selection when a training set is generated by either GSCM or the proposed database.\footnote{For the simulation, NYUSIM was used as GSCM, but all GSCMs could follow the same trend.}
	The $x$-axis represents $\chi=\frac{|\pmb{w}_\text{dnn}^*\pmb{H}\pmb{f}_\text{dnn}|^2}{|\pmb{w}_\text{opt}^*\pmb{H}\pmb{f}_\text{opt}|^2}$ in dB where $\pmb{w}_\text{opt}$ and $\pmb{f}_\text{opt}$ are the beamforming vectors at the RX and the TX determined by exhaustive beam selection, and $\pmb{w}_\text{dnn}$ and $\pmb{f}_\text{dnn}$ denote those determined by the proposed beam selection; and $\pmb{H}$ denotes a channel matrix. We assume the BS has a uniform linear array and tabulate simulation parameters in Fig.~\ref{Figure:DNNresults}a.
	The results show that adopting different channel models can result in different conclusions.
	For example, the proposed beam selection algorithm has only 3-6~dB more gain than random beam selection above the median CDF due to the low accuracy (12.8 percent) when we adopt GSCM, so one may conclude the proposed algorithm does not work.
	Meanwhile, with the proposed database, we achieve reasonable beamforming gain and higher accuracy (45.2 percent), so the DNN-based beam selection algorithm using PDP as an input shows its feasibility.
	The reason for the underestimation from GSCM is that although its PDP has a correlation with angular parameters, it is not fully user-specific.
	Therefore, when we evaluate machine-learning-based algorithms and need to consider user-specific channel parameters, map-based channel models should be employed. Also, the proposed database would be a good and convenient option for training these algorithms.

\section{Conclusion}
This article has provided an overview of map-based mm-Wave channel models and guidelines of the categorization of map-based channel parameters that possess the following: the map-based deterministic channel parameter, the map-based stochastic channel parameter, and the map-based hybrid channel parameter. Map-based models should be utilized to consider the various modeling requirements of applications in the mm-Wave range, which possess short-range communication links, realistic indoor environments, various mobility types, and user-specific channel parameters. In addition, map-based models can support a HW measurement validation at both link and system levels so that it can be treated as a supplementary SW testbed/sounder. It would be efficient to perform a channel measurement campaign through a map-based SW channel sounder before firmly establishing the HW-based measurement procedure.
Finally, we have made public the measurement database of the channel parameters for indoor and outdoor scenarios and for evaluating machine learning algorithms. Through the proposed database, we have concluded that researchers should utilize map-based channel models when they propose machine learning-based algorithms to prevent their underestimation.
Our future work will consist of improving the proposed DNN-based beam selection algorithm.
We will also design map-based V2X and A2X mm-Wave channel models and share their data motivated by a fact that the measurement procedure for such models would be complicated because a vehicular channel sounder for V2X links and an airborne channel sounder for A2X links should be developed.

\ifCLASSOPTIONcaptionsoff
\newpage
\fi

\bibliographystyle{IEEEtran}
\bibliography{RTmmwave_reference_etal}

\begin{thebibliography}{10}
\providecommand{\url}[1]{#1}
\csname url@samestyle\endcsname
\providecommand{\newblock}{\relax}
\providecommand{\bibinfo}[2]{#2}
\providecommand{\BIBentrySTDinterwordspacing}{\spaceskip=0pt\relax}
\providecommand{\BIBentryALTinterwordstretchfactor}{4}
\providecommand{\BIBentryALTinterwordspacing}{\spaceskip=\fontdimen2\font plus
\BIBentryALTinterwordstretchfactor\fontdimen3\font minus
  \fontdimen4\font\relax}
\providecommand{\BIBforeignlanguage}[2]{{%
\expandafter\ifx\csname l@#1\endcsname\relax
\typeout{** WARNING: IEEEtran.bst: No hyphenation pattern has been}%
\typeout{** loaded for the language `#1'. Using the pattern for}%
\typeout{** the default language instead.}%
\else
\language=\csname l@#1\endcsname
\fi
#2}}
\providecommand{\BIBdecl}{\relax}
\BIBdecl

\bibitem{Lim_WCM18}
Y.-G. Lim~\textit{et al.}, ``Waveform multiplexing for new radio: Numerology
  management and {3D} evaluation,'' \emph{{IEEE} Wireless Commun. Mag.},
  vol.~25, no.~5, pp. 86--94, Oct. 2018.

\bibitem{SSCM_TMTT}
M.~K. Samimi and T.~S. Rappaport, ``{3-D} millimeter-wave statistical channel
  model for {5G} wireless system design,'' \emph{{IEEE} Trans. Microw. Theory
  Tech.}, vol.~64, no.~7, pp. 2207--2225, July 2016.

\bibitem{METISmodel}
{ICT-317669 METIS Project deliverable D1.4 v.3}, \emph{{METIS} Channel Models},
  June 2015.

\bibitem{3GPP_mmwave}
{3GPP {TR} 38.900 {V}14.3.0}, \emph{Study on channel model for frequency
  spectrum above 6 GHz}, June 2018.

\bibitem{ComMag_METIS}
J.~Medbo~\textit{et al.}, ``Radio propagation modeling for {5G} mobile and
  wireless communications,'' \emph{{IEEE} Commun. Mag.}, vol.~54, no.~6, pp.
  144--151, June 2016.

\bibitem{Kwon_TMTT}
T.~Kwon~\textit{et al.}, ``{RF} lens-embedded massive {MIMO} systems:
  Fabrication issues and codebook design,'' \emph{{IEEE} Trans. Microw. Theory
  Tech.}, vol.~64, no.~7, pp. 2256--2271, July 2016.

\bibitem{YJ_ComMag18}
Y.-J. Cho~\textit{et al.}, ``{RF} lens-embedded antenna array for mmwave
  {MIMO}: Design and performance,'' \emph{{IEEE} Commun. Mag.}, vol.~56, pp.
  42--48, July 2018.

\bibitem{RTmmwave_JSAC}
B.~Ai~\textit{et al.}, ``On indoor millimeter wave massive {MIMO} channels:
  Measurement and simulation,'' \emph{{IEEE} J. Sel. Areas Commun.}, vol.~35,
  no.~7, pp. 1678--1690, July 2017.

\bibitem{Rappaport_SiteSpecific}
S.~Y. Seidel and T.~S. Rappaport, ``Site-specific propagation prediction for
  wireless in-building personal communication system design,'' \emph{{IEEE}
  Trans. Veh. Technol.}, vol.~43, no.~4, pp. 879--891, Nov 1994.

\bibitem{WiSE_JSAC}
V.~Erceg~\textit{et al.}, ``Comparisons of a computer-based propagation
  prediction tool with experimental data collected in urban microcellular
  environments,'' \emph{{IEEE} J. Sel. Areas Commun.}, vol.~15, no.~4, pp.
  677--684, May 1997.

\bibitem{Klautau_18ITA}
A.~Klautau~\textit{et al.}, ``{5G} {MIMO} data for machine learning:
  Application to beam-selection using deep learning,'' in \emph{Proc. Inf.
  Theory and Appl. Workshop (ITA)}, Feb. 2018, pp. 1--9.

\bibitem{Ju18GC}
S.~Ju and T.~S. Rappaport, ``Millimeter-wave extended {NYUSIM} channel model
  for spatial consistency,'' in \emph{Proc. IEEE Global Commun. Conf.}, Dec.
  2018, pp. 1--6.

\bibitem{RTmmwave_Access}
V.~Degli-Esposti~\textit{et al.}, ``Ray-tracing-based mm-wave beamforming
  assessment,'' \emph{IEEE Access}, vol.~2, pp. 1314--1325, 2014.

\bibitem{ITU_material}
{ITU-R P.2040}, \emph{Effects of building materials and structures on radiowave
  propagation above about {100 MHz}}, Dec. 2013.

\bibitem{Xing18GC}
Y.~Xing and T.~S. Rappaport, ``Propagation measurement system and approach at
  140 {GHz}- moving to {6G} and above {100 GHz},'' in \emph{Proc. IEEE Global
  Commun. Conf.}, Dec. 2018, pp. 1--6.

\end{thebibliography}

\end{document}